\documentclass[aps,twocolumn,prl,superscriptaddress,footinbib]{revtex4-2}
\usepackage{amsmath,amssymb,bm}
\usepackage{graphicx}
\usepackage{epstopdf}
\usepackage{latexsym}
\usepackage[normalem]{ulem} 
\usepackage[usenames,dvipsnames]{color}
\usepackage{hyperref}
\usepackage{natbib}
\usepackage{braket}
\begin{document}

\newcommand{\Rev }[1]{{\color{black}{#1}\normalcolor}} 
\newcommand{\Com}[1]{{\color{black}{#1}\normalcolor}} 

\title{A cavity-QED quantum simulator of dynamical phases of a BCS superconductor}
\date{\today}

\author{Robert J. Lewis-Swan}
\affiliation{Homer L. Dodge Department of Physics and Astronomy, The University of Oklahoma, Norman, Oklahoma 73019, USA}
\affiliation{Center for Quantum Research and Technology, The University of Oklahoma, Norman, Oklahoma 73019, USA}
\affiliation{JILA, NIST, Department of Physics, University of Colorado,  Boulder, CO 80309, USA}
\affiliation{Center for Theory of Quantum Matter, University of Colorado, Boulder, CO 80309, USA}
\author{Diego Barberena}
\affiliation{JILA, NIST, Department of Physics, University of Colorado,  Boulder, CO 80309, USA}
\affiliation{Center for Theory of Quantum Matter, University of Colorado, Boulder, CO 80309, USA}
\author{Julia R.~K. Cline}
\affiliation{JILA, NIST, Department of Physics, University of Colorado,  Boulder, CO 80309, USA}
\author{Dylan J. Young}
\affiliation{JILA, NIST, Department of Physics, University of Colorado,  Boulder, CO 80309, USA}
\author{James K. Thompson}
\affiliation{JILA, NIST, Department of Physics, University of Colorado,  Boulder, CO 80309, USA}
\author{Ana Maria Rey}
\affiliation{JILA, NIST, Department of Physics, University of Colorado,  Boulder, CO 80309, USA}
\affiliation{Center for Theory of Quantum Matter, University of Colorado, Boulder, CO 80309, USA}

\begin{abstract}
We propose to simulate dynamical phases of a BCS superconductor using an ensemble of cold atoms trapped in an optical cavity. Effective Cooper pairs are encoded via internal states of the atoms and attractive interactions are realized via the exchange of virtual photons between atoms coupled to a common cavity mode. Control of the interaction strength combined with a tunable dispersion relation of the effective Cooper pairs allows exploration of the full dynamical phase diagram of the BCS model, as a function of system parameters and the prepared initial state. Our proposal paves the way for the study of non-equilibrium features of quantum magnetism and superconductivity by harnessing atom-light interactions in cold atomic gases.

\end{abstract}

\maketitle  

\noindent{\it Introduction:} The development of a generic framework to understand the properties of non-equilibrium quantum states is a long-standing challenge in modern physics. Theoretical work \cite{Sciolla_2011,Heyl_DPTtheory_2013,Heyl_2018,Silva_DQPT_2018,Morigi2019,Ritsch_2020} combined with technical advances in the control and characterization of many-body physics in cold atom experiments \cite{Roos_DPT_2017,Monroe_DPT_2017,Muniz2020,Thywissen_DPT_2018,Duan_DPT_2020,Chu2020,Baumann2010,Klinder2015,Kroeze2018} has led to new developments in this direction, such as extending the concept of phase transitions to non-equilibrium situations. Specifically, dynamical phase transitions (DPTs) \cite{eckstein09,Schiro10,Sciolla_2011,gambassi,Knap,Dicke2019} have been introduced to classify distinct regimes of dynamical behaviour that arise after a sudden quench of a control parameter in a closed system. DPTs are characterized by the existence of a time-averaged order parameter that demonstrates non-analytic behaviour at the boundary between dynamical phases. 

A long standing example of such dynamical phases are those predicted to emerge from quenches of Bardeen–Cooper–Schrieffer (BCS) superconductors, which has been theoretically investigated in both the condensed matter\cite{Volkov_1974,Spivak_BCS_2004,Altshuler_2005,Enolskii_2005,Altshuler_2006,Levitov_2006,Foster_2015,Hector_2018,Hector_2019,Hector_2020} and high energy communities \cite{Pehlivan2011}. However, experimental progress towards observing these phases has been limited so far to transient dynamics on rapid time-scales in terahertz pump-probe experiments \cite{Shimano_2013,Shimano_2014}. Recent proposals to enhance pairing  by coupling materials to cavities and adjustable external laser driving might facilitate probing the predicted BCS phases in solid state systems \cite{Gao2020}. 

Here, motivated by developments studying dynamical phase transitions in state-of-the-art quantum simulators, we present a proposal to emulate the non-equilibrium dynamics of the BCS model of superconductivity with cavity-QED \cite{Norcia_2018,Muniz2020,Davis_2020,Vaidya_2018,Baumann2010,RevMod2013}. Our scheme leverages the tunability and control available in this platform to map out the dynamical phase diagram over a broad range of system parameters and initial states, demonstrating the power of cavity-QED systems as quantum simulators of superconductivity and quantum magnetism \cite{Strack,Gopalakrishnan,kelly2020,colella2019}.

\noindent{\it BCS model and dynamical phases:} The BCS model of superconductivity for $s$-wave interacting fermions is characterized by the Hamiltonian \cite{Gurarie_2007},
\begin{equation}
    \hat{H} = -\chi \sum_{\mathbf{k},\mathbf{k}^{\prime}} \hat{c}^{\dagger}_{\mathbf{k},\uparrow}\hat{c}^{\dagger}_{-\mathbf{k},\downarrow} \hat{c}_{\mathbf{k}^{\prime},\uparrow}\hat{c}_{-\mathbf{k}^{\prime},\downarrow} 
    + \sum_{\mathbf{k}, \sigma} \varepsilon_{\mathbf{k}} \hat{c}^{\dagger}_{\mathbf{k},\sigma} \hat{c}_{\mathbf{k},\sigma} . \label{eqn:HBCS_fermion}
\end{equation}
Here, $\hat{c}^{\dagger}_{\mathbf{k},\sigma}$ ($\hat{c}_{\mathbf{k},\sigma}$) creates (annihilates) a fermion of momentum $\mathbf{k}$ and spin $\sigma = \uparrow,\downarrow$. The first term describes attractive $s$-wave interactions $\chi \geq 0$ that lead to the formation of Cooper pairs. The single-particle dispersion is $\varepsilon_{\mathbf{k}} = \mathbf{k}^2/(2m) - \mu$ with $\mu$ the chemical potential and $m$ the particle mass. Throughout the manuscript we set $\hbar = 1$. 

\begin{figure}[tb]
    \includegraphics[width=8cm]{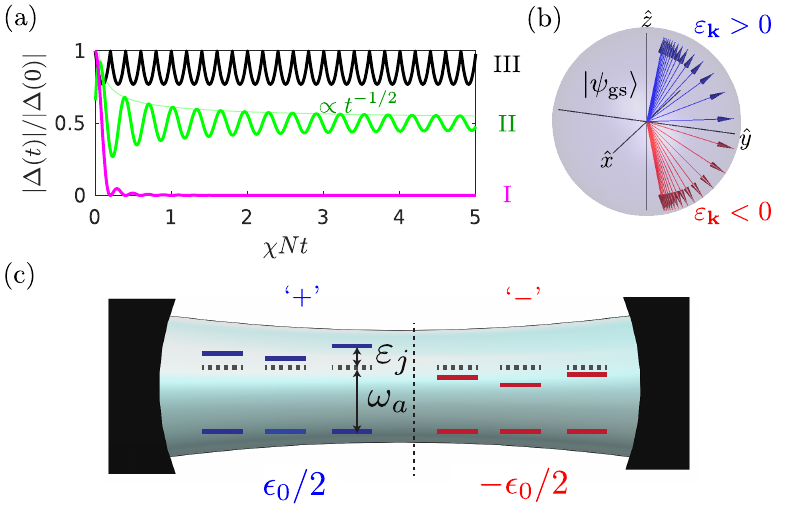}
    \caption{(a) BCS dynamical phases illustrated by the pairing amplitude $\vert \Delta(t) \vert$. Characteristic $t^{-1/2}$ decay of phase II is indicated by the faded line. (b) Example BCS ground-state on the Bloch sphere. The single-particle inversion $\langle \hat{\sigma}^z_{\mathbf{k}} \rangle$ correlates with the sign of the dispersion $\varepsilon_{\mathbf{k}}$ [Eq.~(\ref{eqn:BCS_gs_spin})]. (c) BCS physics can be simulated in a cavity by encoding a spin-$1/2$ into a pair of internal atomic states with transition  frequency $\omega_a$, which are coupled to a single common cavity mode. The spin-$1/2$ atoms are divided into two ensembles (shown as blue and red) featuring mean  energy splittings with opposite sign, $\pm\epsilon_0/2$.}
    \label{fig:CavitySchematic}
\end{figure}

This ``reduced'' BCS model assumes that only Cooper pairs are created and destroyed with zero center-of-mass momentum and neglects pair-breaking
processes, so the low-energy physics can be described using only the presence or absence of Cooper pairs at each momentum mode. The physics of the model is further simplified by introducing Anderson pseudospin-$1/2$ operators,
\begin{equation}
    \hat{\sigma}^-_{\mathbf{k}} = \hat{c}_{\mathbf{k},\uparrow}\hat{c}_{-\mathbf{k},\downarrow}, \quad
    \hat{\sigma}^{z}_{\mathbf{k}} =  \hat{c}^{\dagger}_{\mathbf{k},\uparrow}\hat{c}_{\mathbf{k},\uparrow} + \hat{c}^{\dagger}_{-\mathbf{k},\downarrow}\hat{c}_{-\mathbf{k},\downarrow} -1 .
\end{equation}
The two eigenstates of $\hat{\sigma}^z_{\mathbf{k}}$ encode the presence/absence of a Cooper pair with momentum $\mathbf{k}$, which are created (annihilated) by $\hat{\sigma}^+_{\mathbf{k}}$ ($\hat{\sigma}^-_{\mathbf{k}}$). Equation (\ref{eqn:HBCS_fermion}) then becomes:
\begin{equation}
    \hat{H} = -\chi \sum_{\mathbf{k},\mathbf{k}^{\prime}} \hat{\sigma}^+_{\mathbf{k}}\hat{\sigma}^-_{\mathbf{k}^{\prime}} + \sum_{\mathbf{k}} \varepsilon_{\mathbf{k}} \hat{\sigma}^z_{\mathbf{k}} 
     = -\chi \hat{S}^+\hat{S}^- + \sum_{\mathbf{k}} \varepsilon_{\mathbf{k}} \hat{\sigma}^z_{\mathbf{k}} , \label{eqn:HBCS}
\end{equation}
where $\hat{S}^{\pm} = \sum_{\mathbf{k}} \hat{\sigma}^{\pm}_{\mathbf{k}}$ are collective spin operators. 

The ground-state $\vert \psi \rangle_{\mathrm{gs}}$ of (\ref{eqn:HBCS}) within BCS theory is characterized by the expectations \cite{Levitov_2006} 
\begin{equation}
    \langle \hat{\sigma}^+_{\mathbf{k}} \rangle_{\mathrm{gs}} = \frac{1}{2}\frac{\Delta_{\mathrm{gs}}}{\sqrt{\Delta_{\mathrm{gs}}^2 + \varepsilon^2_{\mathbf{k}}}}, \quad \langle \hat{\sigma}^z_{\mathbf{k}} \rangle_{\mathrm{gs}} = \frac{\varepsilon_{\mathbf{k}}}{\sqrt{\Delta_{\mathrm{gs}}^2 + \varepsilon^2_{\mathbf{k}}}} , \label{eqn:BCS_gs_spin}
\end{equation}
as shown schematically on the Bloch sphere in Fig.~\ref{fig:CavitySchematic}(b). Here, the BCS pairing gap, $\Delta_{\mathrm{gs}} \equiv \chi \langle \hat{S}^- \rangle_{\mathrm{gs}}$, is defined self-consistently.

Prior studies \Rev {in superconductors  and fermionic superfluids} \cite{Foster_2015,Foster2013,Levitov_2006}  have used  the BCS Hamiltonian (\ref{eqn:HBCS})  to describe the gap dynamics after a quench of the pairing gap from the ground-state value $\Delta_{\mathrm{gs}}$ to a final value $\Delta_{\mathrm{f}}$ \cite{Levitov_2006,Foster_2015}. Equation (\ref{eqn:HBCS}) is expected to provide a valid treatment of the gap dynamics on timescales for which pair breaking processes  can be neglected, provided  the quench is done faster than the inverse of the quasiparticle gap.

In the mean-field (classical) limit the dynamics falls into three distinct dynamical phases according to the behaviour of the magnitude of $\vert\Delta(t)\vert = \chi \vert S^-(t)\vert$. Throughout, we adopt the notation $\mathcal{O}(t) \equiv \langle \hat{\mathcal{O}}(t) \rangle$ when making a mean-field approximation, i.e.,  $\langle\hat{\mathcal{O}}_1(t)\hat{\mathcal{{O}}}_2(t)\rangle=\langle\hat{\mathcal{O}}_1(t)\rangle \langle\hat{\mathcal{{O}}}_2(t)\rangle$. As $t\to\infty$ the dynamics are: Phase I) $\vert\Delta(t)\vert \to 0$, Phase II) $\vert \Delta (t) \vert \to \mathrm{const}$ with transient oscillations that decay as  $\propto t^{-1/2}$, or Phase III) $\vert \Delta (t) \vert$ features persistent oscillations. Illustrations of $\vert \Delta(t) \vert$ in each phase are shown in Fig.~\ref{fig:CavitySchematic}(a). We discuss below how these phases arise from a competition between the interactions and the distribution of single-particle splittings $\varepsilon_{\mathbf{k}}$.


\noindent{\it BCS physics in a cavity-QED simulator:} We propose to explore the phase diagram of the BCS model by emulating the Hamiltonian (\ref{eqn:HBCS}) in a cavity. In our proposed scheme, an ensemble of atoms is distributed in a standing wave optical lattice supported by the cavity. Each atom, which we index by the label $j$, encodes a spin-$1/2$ degree of freedom in a pair of stable internal states, $\vert\uparrow \rangle_j$ and $\vert\downarrow\rangle_j$, which map to the presence or absence of a Cooper pair, respectively. The use of the index $j$ compared to the momentum label $\mathbf{k}$ in a real BCS superconductor will be shown to be irrelevant.

Spin-spin interactions $\propto \hat{S}^+\hat{S}^-$ are mediated by the exchange of virtual photons between atoms via a single common cavity mode (at frequency $\omega_c$) far-detuned from the atomic resonance (at frequency $\omega_a$) \cite{Norcia_2018,Muniz2020, Davis_2019,Shankar_2019}. These photon-mediated interactions are analogous to the phonon-mediated interactions in a BCS superconductor. Tunable (inhomogeneous) single-particle energy shifts $\varepsilon_j\hat{\sigma}^z_j$ can be realized via external fields that generate AC Stark or Zeeman shifts of the internal atomic states.

\begin{figure}[tb]
    \includegraphics[width=8cm]{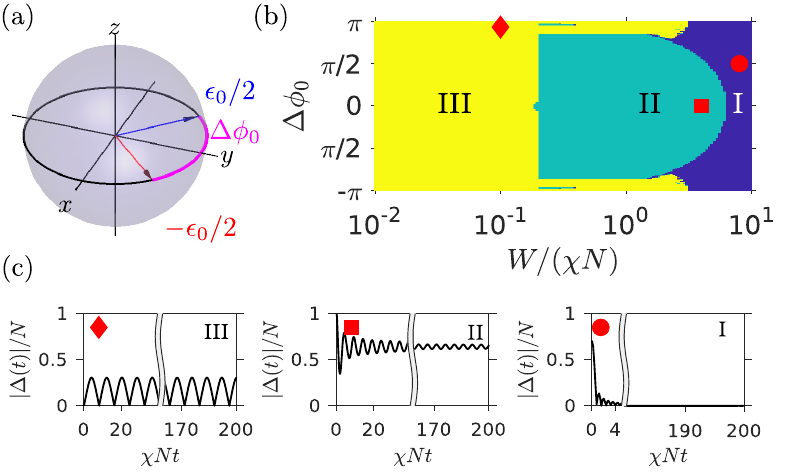}
    \caption{(a) Typical initial state for opening angle $\Delta\phi_0$. The orientation of each ensemble (red and blue collective Bloch vectors) is correlated with the sign of $\pm \epsilon_0$. (b) Mean-field BCS dynamical phase-diagram as a function of $\Delta\phi_0$ andcharacteristic width $W$ of the single-particle noise distribution, with fixed $\epsilon_0/(\chi N) = 0.1$. The phase diagram is evaluated numerically (see Ref.~\cite{SM}) and some small structure (e.g., regions of phase II within phase III) are likely artefacts of the methods precision. (c) Time-traces of the pairing amplitude $\vert\Delta(t)\vert$ for each phase [parameters indicated by marker in (b)].}
    \label{fig:DynamicalPhaseDiagram}
\end{figure}

An important ingredient for the observation of the dynamical phases I-III is the ability to prepare initial states correlated with the distribution of splittings $\varepsilon_j$. For example, in the BCS ground-state [Eq.~(\ref{eqn:BCS_gs_spin})], the sign of the inversion $\langle \hat{\sigma}^z_{\mathbf{k}} \rangle$ of the Anderson pseudospins correlates with the sign of the single-particle dispersion $\varepsilon_{\mathbf{k}}$. Motivated by this case, we consider initial states where the atoms are split into a pair of ensembles where the spin configuration of the atoms in each ensemble is correlated with the sign of the  ensemble's average splitting. Concretely, we consider $2N$ atoms divided into two equal ensembles and initialized as a product of coherent spin-states \cite{Radcliffe_1971} lying on the equatorial plane of the Bloch sphere separated by a relative azimuthal opening angle $\Delta\phi_0$: $\vert \psi_0 \rangle = \vert \pi/2,\Delta\phi_0/2\rangle_+ \otimes \vert \pi/2,-\Delta\phi_0/2\rangle_-$ [see Fig.~\ref{fig:CavitySchematic} and Fig.~\ref{fig:DynamicalPhaseDiagram}(a)] where the subscript $\pm$ denotes each ensemble. Here, $\vert \theta, \phi \rangle \equiv \bigotimes_{j} \left[ \mathrm{cos}(\theta/2)\vert \downarrow \rangle_j + e^{ i\phi}\mathrm{sin}(\theta/2) \vert \uparrow \rangle_j \right]$ where the product runs over $j=1,...,N$ or $j=N+1,...,2N$ atoms respectively for the $\pm$ ensembles. Lastly, following the BCS ground-state we assume a uniform distribution of splittings $\varepsilon_j \in [\pm\epsilon_0/2 - W/4, \pm\epsilon_0/2 + W/4]$ where the sign of $\epsilon_0$ differs for each ensemble and is matched to the sign of $\pm\Delta\phi_0/2$. It is the mean $\pm\epsilon_0/2$ and characteristic width $W/2$ rather than the precise distribution of $\varepsilon_j$ (e.g., uniform or normal) that is important to characterize the physics discussed below.

Preparation of the two ensembles and correlation with $\pm\epsilon_0$ can be achieved by spatially selective energy shifts of atoms in the cavity \cite{Davis_2019} [Fig.~\ref{fig:CavitySchematic}(c)] or by addressing different internal levels \cite{RLS_2018,Norcia_2018} (see later discussion) \footnote{It should also be possible to prepare initial states that are split by their projection along the $z$ direction, rather than the projection along $x$ (set by $\Delta\phi_0$) that would follow even more closely the BCS ground state. However, varying the relative azimuthal opening angle gives similar physics and is more robust to typical experimental constraints \cite{SM}}.

\noindent{\it Accessible dynamical phase diagram:} In Fig.~\ref{fig:DynamicalPhaseDiagram} we explore the accessible dynamical phases. Panel (b) shows the dynamical phase diagram for mean splitting $\epsilon_0/(\chi N) = 0.1$. The phase diagram is computed via a Lax analysis \cite{Enolskii_2005,Levitov_2006,Foster_2015} that is a method for integrable  models, such as Eq.~(\ref{eqn:HBCS}), to determine the frequency spectrum that rules the dynamics of the order parameter. The spectrum is extracted from the roots of  $L^2(u)$, the squared norm of the Lax vector  $\mathbf{L}(u)$,  a polynomial defined in terms  of a  complex variable $u$   that encodes the conserved quantities of the model.  A spectrum with all  real roots  defines phase I, with one pair of complex roots  phase II and  with two pairs of complex roots phase III.  The asymptotic behaviour of $\vert \Delta(t) \vert$ follows from the nature of the roots of $L^2(u)$, which we compute numerically \cite{SM}.

Physically, the dynamical phases depend on the competition between single-particle dephasing generated by $W$ and $\epsilon_0$ and the spin-locking effect generated by the interactions  with strength set by  $\chi N$ \cite{Laloe1982,Gully_1984,Johnson_1984,Bashkin1986,McGuirk2002,Du2008,Deutsch2010,Kleine2011}. The spin-locking is induced by the existence of a many-body gap that suppresses local spin flips and favors spin alignment \cite{Norcia_2018,Davis_2020,Rey_2008}. Such behaviour also resembles synchronization observed in arrays of coupled oscillators with dissipation \cite{Zhu2015,kuramoto}. Other consequences associated with the many-body gap include the stabilization of localization effects in fully connected models under specific initial conditions \cite{Santos2016,Celardo2016}.

\begin{figure}[tb]
    \includegraphics[width=8cm]{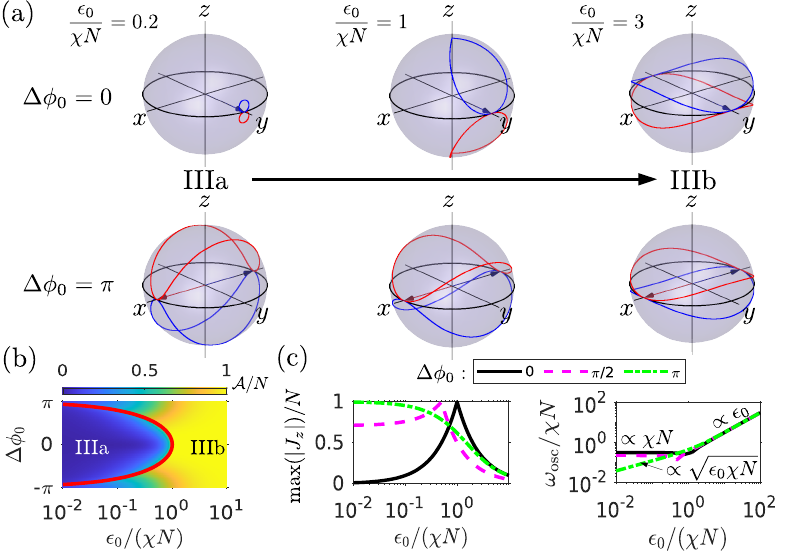}
    \caption{(a) Typical trajectories of the collective Bloch vector of each ensemble (red and blue) for $W \ll \chi N, \epsilon_0$ as $\epsilon_0$ is tuned between phases IIIa and IIIb. (b) Phase diagram characterized by amplitude $\mathcal{A} \equiv \mathrm{max}(\vert \Delta(t) \vert) - \mathrm{min}(\vert \Delta(t) \vert)$ of oscillations in $\vert \Delta(t) \vert$. The critical boundary  $\epsilon_0^{\mathrm{c}}$ between phases IIIa and IIIb is indicated by the red line. (c) Maximum of the total inversion difference $J_z = \left(\sum_{j \in +} \sigma^z_j - \sum_{j \in -} \sigma^z_j\right)/2$ and frequency, $\omega_{\mathrm{osc}}$, of oscillations of $\vert \Delta(t) \vert$ as a function of $\epsilon_0$.}
    \label{fig:SubPhase3}
\end{figure}

For small inhomogeneity, $W \ll \epsilon_0, \chi N$ we predict phase III dynamics independent of the opening angle $\Delta\phi_0$. Within each ensemble a gap opens between the manifold of collective states (this includes the initial fully polarized states) and those that are spatially inhomogeneous, preventing dephasing of the individual spins of each ensemble. In addition, the interplay between the homogeneous single-particle energy splitting $\pm \epsilon_0$ (that generates precession of the ensembles in opposing directions about the $z$-axis of the Bloch sphere), and the collective interaction (that also drives a rotation of each ensemble  along a common self-generated axis set by the total transverse magnetization \cite{Norcia_2018,RLS_2018}), leads to persistent nonlinear oscillations in the effective pairing amplitude $\vert \Delta(t) \vert = \vert S^+(t) \vert$ [see also Fig.~\ref{fig:SubPhase3}(a)]. 

Phase II emerges for $2\epsilon_0< W \lesssim \chi N$ and opening angles away from $\Delta \phi \approx \pm\pi$ \cite{SM}. The transition from phase III to II is driven by the ensembles no longer having a well-defined relative energy splitting correlated with their initial orientation. Thus, in contrast to phase III, spin locking of the \emph{entire} ensemble of $2N$ atoms determines the dynamics. This means that while the pairing amplitude $\vert\Delta(t)\vert$ remains large, oscillations are transient and suppressed rather than stabilized by the interactions.

Finally, phase I emerges for $W \gtrsim \chi N$ independent of $\epsilon_0$. Single-particle physics dominates for all initial conditions and the pairing amplitude vanishes due to rapid dephasing of the individual spins, $\vert \Delta(t) \vert \to 0$.

Beyond these three known regimes we also predict the emergence of two previously unidentified sub-phases within phase III for $W \ll \chi N, \epsilon_0$, which we label as IIIa and IIIb. These sub-phases are delineated by a critical splitting $\epsilon_0^{\mathrm{c}} = \frac{\chi N}{2}[1 + \mathrm{cos}(\Delta\phi_0)]$ \cite{SM}. Phase IIIa), $\epsilon_0 < \epsilon_0^{\mathrm{c}}$, is dominated by interactions and characterized by a strictly non-zero pairing amplitude, $\vert \Delta(t) \vert > 0$ that exhibits nonlinear oscillations with an approximate frequency $\omega_{\mathrm{osc}}\propto \chi N$. Phase IIIb) is characterized by the pairing amplitude periodically vanishing, $\vert \Delta(t) \vert = 0$, and the physics is dominated by the single-particle splitting $\epsilon_0$ such that the frequency of oscillations scales as $\omega_{\mathrm{osc}}\propto \epsilon_0$.

In Fig.~\ref{fig:SubPhase3}(a) we illustrate typical trajectories of the collective Bloch vector of each ensemble in sub-phases IIIa and IIIb for an initial state with $\Delta\phi_0 = 0$, which are representative of the dominant physics for $\vert\Delta\phi_0\vert \lesssim \pi$. For small $\epsilon_0 \ll \epsilon_0^{\mathrm{c}}$ and $\Delta\phi_0 \ll \pi$ the Bloch vectors remain trapped close to their initial polarization due to the strong interactions, leading to $\vert \Delta(t) \vert > 0$. As $\epsilon_0$ increases nearer to the transition, $\epsilon_0^{\mathrm{c}}$, the interactions still dominate and their interplay with the single particle term leads to a deflection of the trajectories of the Bloch vectors close to the north and south poles. Above $\epsilon_0^{\mathrm{c}}$ the trajectories abruptly snap to large orbits near the equator and quickly approach the precession expected for two independent ensembles (e.g., dominated by the $\hat{\sigma}_z$ term of the Hamiltonian). Even though phase IIIb is technically absent for $\Delta\phi_0 = \pm \pi$ by our definition (as $\vert \Delta(0) \vert = 0$), we still observe rich non-trivial oscillations for $\epsilon_0 \ll \chi N$ with frequency $\omega_{\mathrm{osc}}\propto \sqrt{\epsilon_0\chi N}$ \cite{SM}.

Quantitatively, the IIIa and IIIb sub-phases are delineated by abrupt changes in different observables, including the magnitude $\mathcal{A} \equiv \mathrm{max}(\vert \Delta(t) \vert) - \mathrm{min}(\vert \Delta(t) \vert)$ , the frequency $\omega_{\mathrm{osc}}$ of oscillations of $\vert \Delta(t) \vert$ [Figs.~\ref{fig:SubPhase3}(b) and (d)] and the maximum excursion of the collective spins away from the equator of the Bloch sphere, measured by the differential inversion $J_z = \left(\sum_{j \in +} \sigma^z_j - \sum_{j \in -} \sigma^z_j\right)/2$ [Fig.~\ref{fig:SubPhase3}(c)].

\noindent{\it Experimental realization and robustness of proposal:} In a cavity-QED experiment, the dynamical phases can be characterized by detection of intracavity light leaking out through the cavity mirrors \cite{Norcia_2018}. By operating in the limit where the cavity mode is far off-resonance from the atomic transition, the virtual photons that mediate the interactions are adiabatically eliminated and slaved to the spins, such that atomic information is imprinted onto the phase and amplitude of the cavity field via the approximate relation $a(t) \propto S^-(t) \propto \Delta(t)$ \cite{SM,Norcia_2018}. The light intensity then serves as a proxy for the BCS pairing amplitude, $\vert a(t) \vert^2 \propto \vert \Delta(t)\vert^2$, while the frequency spectrum of $a(t)$ can also be a useful diagnostic to distinguish the dynamical phases. Moreover, by continuously performing heterodyne detection of the small amount of light leaking through the cavity mirrors we are able to in principle construct time-traces of the pairing amplitude within a single experimental trial. 

To demonstrate our proposal is robust to relevant decoherence and technical factors, we model an experiment where the spin-$1/2$ is encoded using the narrow linewidth $^{1}$S$_0$-$^{3}$P$_1$ optical transition of $^{87,88}$Sr. Here, sub-ensembles can be prepared via spatially dependent light-shifts from the side of the cavity, or in $^{87}$Sr by applying spatially dependent magnetic fields and addressing the $\pm9/2$ nuclear spin levels of the transition. We use parameters from Ref.~\cite{Muniz2020} and include single-particle decoherence due to the natural linewidth of the transition $\gamma/(2\pi) = 7.5$~kHz and spatially inhomogeneous atom-light coupling arising due to the incommensurate wavelengths of the standing wave optical lattice confining the atoms and the relevant  cavity mode \cite{Muniz2020,Norcia_2018,SM}. The latter leads to a spatial modulation of the spin-spin interactions $\chi\to \chi_{i,j}$.  Our predictions should also be qualitatively relevant for other cavity-based systems that can realize an effective $\chi\hat{S}^+\hat{S}^-$ interaction, e.g., Raman transitions \cite{Davis_2019,Shankar_2019}.


\begin{figure}[!]
    \includegraphics[width=8cm]{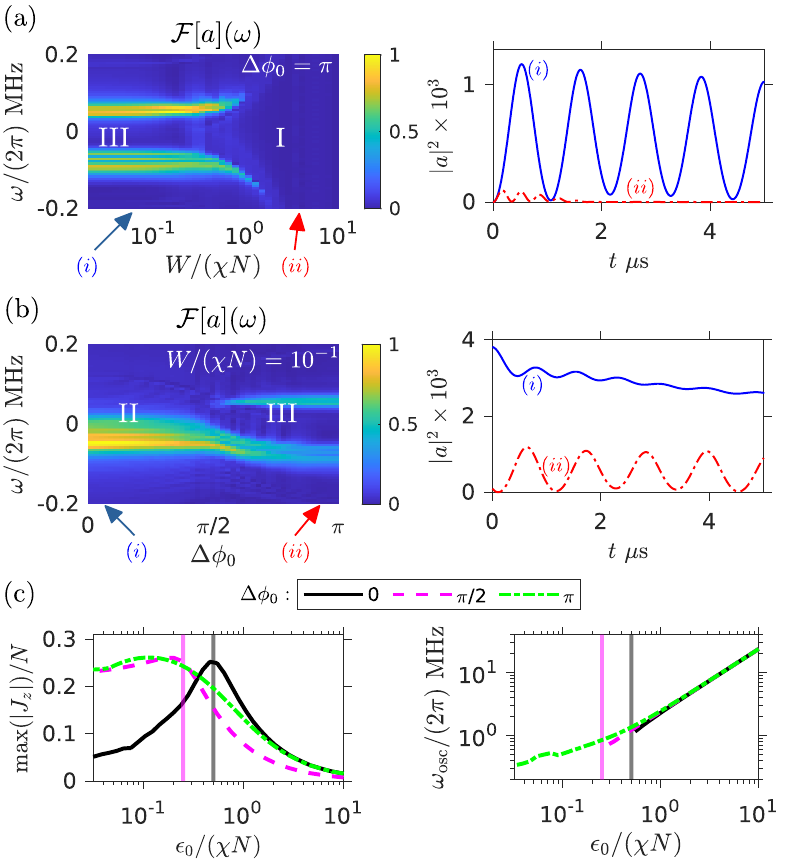}
    \caption{Dynamics of intracavity field $a(t)$. (a) Frequency spectrum of intracavity field, $\mathcal{F}[a](\omega)$, as a function of  $W/(\chi N)$, and typical timetraces of $\vert a \vert^2$ in phases (i) III and (ii) I. Initial state is $\Delta\phi_0 = \pi/2$. (b) Same, but as a function of opening angle $\Delta\phi_0$ and time traces are in phases (i) II and (ii) III. Simulations are for fixed $W/(\chi N) = 0.1$. Both (a) and (b) use fixed $\epsilon_0/(\chi N) = 0.1$ and color scales are normalized. (c) Signatures of phase IIIa and phase IIIb in differential inversion $J_z$ and oscillation frequency $\omega_{\mathrm{osc}}$ of  $\vert a \vert^2$ for $W = 0$ for different $\Delta\phi_0$. Critical  $\epsilon_0^{\mathrm{c}}$ for each $\Delta\phi_0$ is indicated by a vertical line. The absence of plotted results for $\omega_{\mathrm{osc}}$ below the approximate transition $\epsilon_0^{\mathrm{c}}$ for each $\Delta\phi_0$ indicates the lack of appreciable oscillations in the simulations. All relevant parameters (e.g., $g,\gamma$ and $\kappa$) are taken from Refs.~\cite{Muniz2020,Norcia_2018} and results are rescaled for $N = 10^6$ (see also Ref.~\cite{SM}).}
    \label{fig:ExpPrediction}
\end{figure}

In Fig.~\ref{fig:ExpPrediction}(a) we model the transition between phases I and III as a function of the inhomogeneity strength $W$ at fixed $\epsilon_0/(\chi N) = 0.1$ and initial state $\Delta\phi_0 = \pi$. The phases are distinguished in the frequency spectrum of the cavity field $\mathcal{F}[a](\omega)$, with phase III signaled by a pair of robust peaks in the spectrum that disappear in phase I. The peaks are consistent with the entwined but distinguishable precession of the two ensembles that leads to beating of the intensity $\vert a(t) \vert^2$ as shown in the accompanying time-trace. The oscillations in the intracavity intensity $\vert a(t) \vert^2$ are robust to the inhomogeneous interactions and the exponential decay induced by $\gamma$. The transition between the phases occurs at  $W/(\chi N) \approx \pi/2$, which is consistent with the model Eq. (\ref{eqn:HBCS}) when the inhomogeneous atom-light coupling is taken into account by a simple rescaling to the corresponding mean value  $\chi\to\overline{\chi_{i,j}}=\chi/2$ \cite{SM}.

Similarly, the phase II-III transition can be observed by varying the initial opening angle $\Delta\phi_0$ at fixed $W/(\chi N) = \epsilon_0/(\chi N) = 0.1$. The spectrum of the intracavity field shows the signature dual peaks of phase III for $\pi/2 \lesssim \Delta\phi_0 \leq \pi$, while phase II is signaled by a single peak for $0 \leq \Delta\phi_0 \lesssim \pi/2$. The latter indicates dynamics of a single collective ensemble, with finite but non-oscillatory pairing amplitude. 

Lastly, signatures of the phase IIIa-IIIb transition in the differential inversion and oscillation frequency of $\vert a(t) \vert^2$ are shown in Fig.~\ref{fig:ExpPrediction}(c). Decoherence blunts the expected cusp in the inversion, although the peak value lines up closely with the expected transition upon accounting for inhomogeneous interactions. The oscillation frequency clearly distinguishes the trivial/non-trivial regimes for $\Delta\phi_0 = \pi$. On the other hand, we find that for $\Delta\phi_0 = 0,\pi/2$ the relatively small oscillations in $\vert a(t) \vert^2$ predicted for phase IIIa are destroyed by decoherence,and instead the transition between IIIa and IIIb is marked by an abrupt vanishing of any discernible peak in the spectrum (indicated by the absence of data).

\noindent{\it Conclusions:} We have reported a proposal to observe the dynamical phases of a BCS superconductor in a cavity-QED quantum simulator. Realizing these phases via a spin degree of freedom instead of actual Cooper pairs overcomes the need to reach the ultra cold temperatures at which pairing occurs. The versatility of this platform allows us to probe the dependence of the dynamical phases on the initial state and system parameters in a controllable, isolated setting. Our predictions pave the way for future studies of more complex non-equilibrium phenomena in models of quantum magnetism and superconductivity so far not seen in real materials or high energy systems.

\begin{acknowledgments}
\noindent{\textit{Acknowledgements:}} We acknowledge helpful discussions with Anjun Chu, Nathan Schine, Victor Gurarie and Emil Yuzbashyan. This work is supported by the AFOSR grant FA9550-18-1-0319, by the DARPA and ARO grant W911NF-16-1-0576, the ARO single investigator award W911NF-19-1-0210, the NSF PHY1820885, NSF JILA-PFC PHY-1734006 and NSF QLCI-2016244 grants, and by NIST.
\end{acknowledgments}

\bibliography{library}

\newpage

\onecolumngrid
\vspace{\columnsep}
\begin{center}
\textbf{\large Supplemental Material: A cavity-QED quantum simulator of dynamical phases of a BCS superconductor}
\end{center}
\vspace{\columnsep}
\twocolumngrid

\setcounter{equation}{0}
\setcounter{figure}{0}
\setcounter{table}{0}
\setcounter{page}{1}
\makeatletter
\renewcommand{\theequation}{S\arabic{equation}}
\renewcommand{\thefigure}{S\arabic{figure}}

\section{Lax analysis}
We gain analytic insight into the dynamical phase diagram of the BCS model by employing a Lax analysis. This is a a tool that, building on the integrability of the BCS model, allows for the characterization and even solution of the asymptotic ($t\to\infty$) dynamics. In the following sections we give a brief summary of results in the classical limit, pertinent in particular for Figs.~2 and 3 of the main text. A more detailed discussion of the Lax analysis can be found in, e.g., Ref.~\cite{Enolskii_2005,Foster_2015} and references therein.

\subsection{Lax vector and classification of dynamical phase diagram}
Throughout our analysis we work with the pseudospin BCS Hamiltonian [Eq.~(3) of the main text],
\begin{equation}
    \hat{H} = \chi \hat{S}^+\hat{S}^- + \sum_{j} \varepsilon_{j} \hat{\sigma}^z_{j} . \label{eqn:BCSham}
\end{equation}
We will focus on the mean-field (classical) dynamics of the model generated by the approximation $\langle \hat{\mathcal{O}}_1 \hat{\mathcal{O}}_2 \rangle = \langle \hat{\mathcal{O}}_1 \rangle \langle \hat{\mathcal{O}}_2 \rangle$ and adopt the notation $\mathcal{O} \equiv \langle \hat{\mathcal{O}} \rangle$ for simplicity. 

Given the form of the Hamiltonian (\ref{eqn:BCSham}) we also introduce the associated (mean-field) Lax vector \cite{Enolskii_2005,Levitov_2006,Foster_2015},
\begin{equation}
    \vec{L}(u) = \frac{1}{2}\sum_j \frac{\vec{\sigma}_j(0)}{u - \varepsilon_j} - \frac{\hat{z}}{\chi} . \label{eqn:LaxDefn}
\end{equation}
The Lax vector is explicitly defined with respect to the initial state characterized by the expectation values $\vec{\sigma}_j(0) = (\langle \hat{\sigma}^x_j(0) \rangle, \langle \hat{\sigma}^y_j(0) \rangle, \langle \hat{\sigma}^z_j(0) \rangle)$.

The mean-field dynamical phase diagram of $\hat{H}$ given the initial conditions $\vec{\sigma}_j(0)$ can be constructed by an analysis of the properties of $\vec{L}(u)$ \cite{Enolskii_2005,Levitov_2006,Foster_2015}. Specifically, the dynamical phases are defined in terms of the properties of the \emph{complex} roots $\{u_1,u_2,...,u_n\}$ of the equation $\vec{L}(u)\cdot\vec{L}(u) = 0$: Phase I corresponds to zero complex roots, phase II is defined by a single pair of complex roots, and phase III is accompanied by two pairs of complex roots. 

To be concrete in our analysis we must specify the initial states which are fed into the definition of the Lax vector. In the main text we consider splitting the atoms inte a pair of ensembles, each of $N$ atoms, initialized as a product of coherent spin-states \cite{Radcliffe_1971} lying on the equatorial plane of the Bloch sphere separated by a relative azimuthal opening angle $\Delta\phi_0$: $\vert \psi_0 \rangle = \vert \pi/2,\Delta\phi_0/2\rangle_+ \otimes \vert \pi/2,-\Delta\phi_0/2\rangle_-$ (see Fig.~1 and Fig.~2(a) of the main text) where the subscript $\pm$ denotes each ensemble. Here, we have used $\vert \theta, \phi \rangle \equiv \bigotimes_{j} \left[ \mathrm{cos}(\theta/2)\vert \downarrow \rangle_j + e^{ i\phi}\mathrm{sin}(\theta/2) \vert \uparrow \rangle_j \right]$ where the product over $j$ runs over $j=1,...,N$ or $j=N+1,...,2N$ atoms respectively for the $\pm$ ensembles. For $\vert \psi_0 \rangle$ we then have $\vec{\sigma}_j(0) = (\mathrm{cos}(\Delta\phi_0/2), \pm\mathrm{sin}(\Delta\phi_0/2),0)$ where the $\pm$ is correlated with the ensemble. Moreover, we assume the single-particle energies are sampled uniformly from $\varepsilon_j \in [\pm \epsilon_0 - W/2, \pm \epsilon_n + W/2]/2$ for each ensemble. Substituting the state and single-particle dispersion into Eq.~(\ref{eqn:LaxDefn}) and taking the continuum limit as $N\to\infty$ we compute the Lax vector $\vec{L}(u) = L_x(u)\hat{x} + L_y(u)\hat{y} - (1/\chi)\hat{z}$,
\begin{widetext}
\begin{equation}
    \begin{gathered}
    \chi L_x(u) = \frac{2\chi N}{W}\mathrm{cos}\left( \frac{\Delta\phi_0}{2} \right) \left[ \mathrm{ArcTanh}\left( \frac{4u}{W - 2\epsilon_0} \right) + \mathrm{ArcTanh}\left( \frac{4u}{W + 2\epsilon_0} \right) \right] , \\
    \chi L_y(u) = \frac{\chi N}{W}\mathrm{sin}\left( \frac{\Delta\phi_0}{2} \right) \mathrm{log} \left[ \frac{(4u)^2 - (W - 2\epsilon_0)^2}{(4u)^2 - (W + 2\epsilon_0)^2} \right] .
    \end{gathered}
\end{equation}
\end{widetext}

Analytically solving for the roots of $\vec{L}(u)\cdot\vec{L}(u) = 0$ is in general not possible for arbitrary choices of system parameters $W, \epsilon_0, \chi N$ and initial state $\Delta\phi_0$. As a consequence we typically solve for the roots using a numerical algorithm, which will be detailed momentarily. However, there are three limiting cases of the parameters for which relatively simple analytic forms of the roots exist and significant insight into the phase diagram can be garnered: (i) $\Delta\phi_0 = \pi$, (ii) $W \ll \chi N, \epsilon_0$, and (iii) $\epsilon_0 \ll W, \chi N$.

For case (i), $\Delta\phi_0 = \pm\pi$, we are able to analytically solve the roots and characterize the phase diagram for any values of $W/(\chi N)$ and $\epsilon_0/(\chi N)$. A straightforward solution of $\vec{L}(u)\cdot\vec{L}(u) = 0$ yields two pairs of complex roots,
\begin{eqnarray}
    u_{1\pm} & = & \pm \frac{1}{4}\sqrt{ \frac{(W-2\epsilon_0)^2 - e^{-i\frac{W}{\chi N}}(W + 2\epsilon_0)^2 }{ 1 - e^{-i\frac{W}{\chi N}} } } , \notag \\
    u_{2\pm} & = & \pm \frac{1}{4}\sqrt{ \frac{(W-2\epsilon_0)^2 - e^{i\frac{W}{\chi N}}(W + 2\epsilon_0)^2 }{ 1 - e^{i\frac{W}{\chi N}} } } , \label{eqn:RootsBB}
\end{eqnarray}
which exist only for $W/(\chi N) < \pi$. If $W/(\chi N) \geq \pi$ then no complex roots exist. This allows us to diagnose the dynamical phases: For $W/(\chi N) < \pi$ the long-time dynamics are phase III, while for $W/(\chi N) \geq \pi$ the dynamics are phase I. Phase II does not exist for any choice of $W/(\chi N)$ or $\epsilon_0/(\chi N)$. 

Case (ii) describes a scenario where the inhomogeneity is small compared to both the interactions and the uniform energy splitting, $W \ll \chi N, \epsilon_0$, and admits an approximate analytic solution for any $\Delta\phi_0$. To compute the roots we first expand the squared Lax vector to lowest-order in $W$,
\begin{eqnarray}
    \vec{L}(u)\cdot\vec{L}(u) & = & 1 + \frac{4\chi^2N^2 }{(4u^2 - \epsilon_0^2)^2} \left[4u^2\mathrm{cos}^2\left( \frac{\Delta\phi_0}{2} \right) \right. \notag \\
    & & \left. + \epsilon_0^2\mathrm{sin}^2\left( \frac{\Delta\phi_0}{2} \right)\right] + \mathcal{O}(W^2) ,
\end{eqnarray}
and then solve $\vec{L}(u)\cdot\vec{L}(u) = 0$ to find 
\begin{widetext}
\begin{equation}
    \begin{gathered}
    u_{1\pm} = \frac{1}{2} \sqrt{ \epsilon_0^2 - \chi^2N^2[1 + \mathrm{cos}(\Delta\phi_0)] \pm \frac{\chi N}{\sqrt{2}}\sqrt{\chi^2N^2[3 + 4\mathrm{cos}(\Delta\phi_0) + \mathrm{cos}(2\Delta\phi_0)] - 8\epsilon_0^2} } , \\
    u_{2\pm} = -\frac{1}{2} \sqrt{ \epsilon_0^2 - \chi^2N^2[1 + \mathrm{cos}(\Delta\phi_0)] \pm \frac{\chi N}{\sqrt{2}}\sqrt{\chi^2N^2[3 + 4\mathrm{cos}(\Delta\phi_0) + \mathrm{cos}(2\Delta\phi_0)] - 8\epsilon_0^2} } . \label{eqn:Roots_SmallW}
    \end{gathered}
\end{equation}
\end{widetext}
The existence of these two pairs of complex roots is insensitive to the choice of $\Delta\phi_0$ and always true for $W \ll \chi N, \epsilon_0$. Thus, we predict that for small inhomogeneity the long-time dynamics is always phase III. 

Lastly, case (iii) considers the case where only the inhomogeneous splitting and interactions dominate the physics, $\epsilon_0 \ll W, \chi N$. Computing the squared Lax vector to lowest-order in $\epsilon_0$,
\begin{equation}
    \vec{L}(u)\cdot\vec{L}(u) = 1 + 16\frac{\chi^2N^2}{W^2}\mathrm{ArcTanh}^2\left( \frac{4u}{W} \right)\mathrm{cos}^2\left(\frac{\Delta\phi_0}{2}\right) ,
\end{equation}
yields a single pair of complex roots 
\begin{equation}
    u_{\pm} = \pm\frac{iW}{4} \mathrm{tan}\left[ \frac{W}{4\chi N}\mathrm{sec}\left(\frac{\Delta\phi_0}{2}\right)\right] ,
\end{equation}
only for $W/(\chi N) < 2\pi \mathrm{cos}(\Delta\phi_0/2)$. If $W/(\chi N) \geq 2\pi \mathrm{cos}(\Delta\phi_0/2)$ no complex roots exist. This allows us to diagnose: For $W/(\chi N) < 2\pi \mathrm{cos}(\Delta\phi_0/2)$ the dynamics is phase II, while for $W/(\chi N) \geq 2\pi \mathrm{cos}(\Delta\phi_0/2)$ the dynamics is phase I. We stress that these phases and the transition at $W/(\chi N) = 2\pi \mathrm{cos}(\Delta\phi_0/2)$ only exist in the limit $\epsilon_0 \ll W, \chi N$. 

In general, we resort to a numerical search for roots of $\vec{L}(u)\cdot\vec{L}(u) = 0$. This is the procedure used to generate Fig.~2a of the main text. We use the inbuilt function \emph{fsolve} of MATLAB 2020a to search for complex-valued roots of the squared Lax vector as a function of system parameters and initial state. An exhaustive search is performed by running many iterations of the root finding algorithm for each choice of $W$, $\epsilon_0$, $\chi N$ and $\Delta\phi_0$. To make this computation efficient we perform our search at fixed $\Delta\phi_0$, $\epsilon_0$, $\chi N$ and begin at small inhomogeneity $W \ll \chi N, \epsilon_0$ for which we can use the results of Eq.~(\ref{eqn:Roots_SmallW}) as the initial guess for the algorithm. The roots are expected to change relatively smoothly as a function of $W$ (until they vanish) and so we use the roots found for the current value of $W$ as the subsequent starting point for the next value of $W$. 

To validate our numerical results, we compare the roots found by the numerical search algorithm to the analytic results for cases (i)-(iii) discussed above in Fig.~\ref{fig:AnalyticRoots}. For case (i), e.g., restricting to $\Delta\phi = \pi$, we observe exact quantitative agreement in terms of both the predicted roots and the presence of dynamical phases I and III (Fig.~\ref{fig:AnalyticRoots}a). More generically, our numerical results for arbitrary $\Delta\phi$ (Fig.~\ref{fig:AnalyticRoots}b) are consistent with the expected result of phase III for $W \ll \epsilon_0, \chi N$ [case ii)] and the boundary between phases II and I at $W/(\chi N) = 2\pi\mathrm{cos}(\Delta\phi_0/2)$ [case iii)]. The breakdown of the latter prediction near $\Delta\phi \approx \pm\pi$ in the numerical results is also consistent, as our assumption that $\epsilon_0 \ll W$ would be in contradiction with the predicted phase boundary.

\begin{figure}[!]
    \includegraphics[width=8cm]{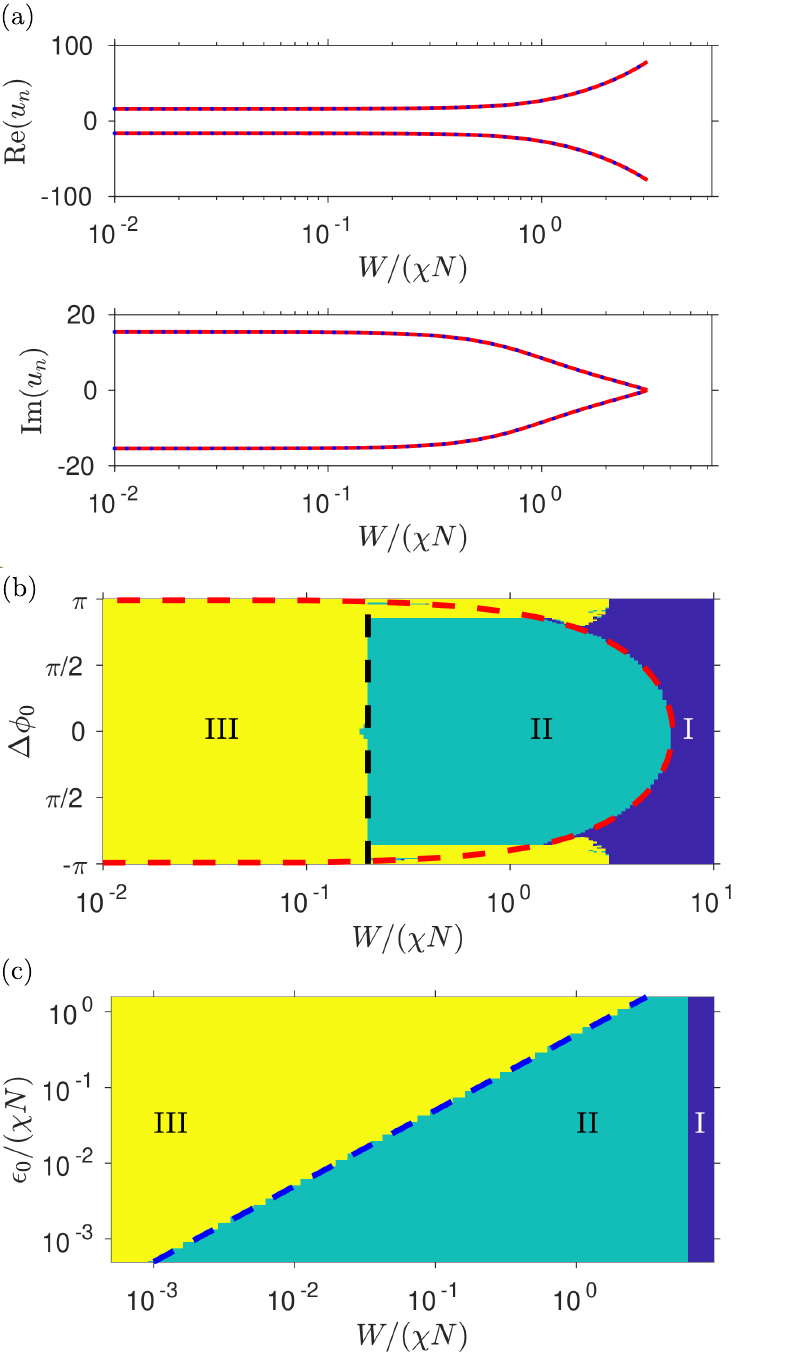}
    \caption{(a) Complex roots of the squared Lax vector, $\vec{L}(u)\cdot\vec{L}(u) = 0$, for $\Delta\phi_0 = \pi$ and $\epsilon_0/(\chi N) = 0.1$. We compare the results of the numerical search algorithm (blue markers) and the predictions of Eq.~(\ref{eqn:RootsBB}) (red lines). No complex roots exist for $W/(\chi N) \geq \pi$. (b) Numerically evaluated dynamical phase diagram as a function of $\Delta\phi_0$ and $W/(\chi N)$ at fixed $\epsilon_0/(\chi N) = 0.1$. We compare the numerical results to analytic predictions for the phase II/I boundary (red dashed line). The phase III/II transition is also indicated at $W=2\epsilon_0$ (dashed black line). (c) Generic phase III/II transition for arbitrary splittings and fixed $\Delta\phi_0 = 0$. We indicate $W=2\epsilon_0$ (dashed blue line) to guide the eye. }
    \label{fig:AnalyticRoots}
\end{figure}

In Fig.~\ref{fig:AnalyticRoots}b (also Fig.~2b of the main text) we identify that a transition between phases III and II occurs when the inhomogeneous and uniform splittings become comparable, $W = 2\epsilon_0 \lesssim \chi N$, and $\Delta\phi \neq \pm\pi$. To support the generality of this observation we compute the phase diagram as a function of both $W/(\chi N)$ and $\epsilon_0/(\chi N)$ for fixed $\Delta\phi_0 =0$ and present the results in Fig.~\ref{fig:AnalyticRoots}c. Our data illustrates that there is always a consistent transition between phases II and III at $W = 2\epsilon_0$.


\subsection{Dynamical phase diagram of alternative initial states}
Throughout the main text we focused on the accessible dynamical phase diagram of the BCS model in the context for specific set of initial conditions parameterized by the state $\vert \psi_0 \rangle = \vert \pi/2,\Delta\phi_0/2\rangle_+ \otimes \vert \pi/2,-\Delta\phi_0/2\rangle_-$ with $\Delta\phi_0 \in [-\pi,\pi]$. Here, $\vert \theta, \phi \rangle_{\pm} \equiv \bigotimes_{j \in \pm } \left[ \mathrm{cos}(\theta/2)\vert \downarrow \rangle_j + e^{ i\phi}\mathrm{sin}(\theta/2) \vert \uparrow \rangle_j \right]$ is a spin coherent state. This choice of state was motivated by both the fact that it shares qualitative features with the BCS ground-state and that it can be prepared relatively accurate in the experiment when taking into account all technical considerations, including inhomogeneity of the atom-light coupling (see Sec.~\ref{sec:ExperimentalModel} and later discussion). The former connection is of note because prior work in the literature studying the BCS dynamical phase diagram has focused on quenches from the BCS ground-state. 

As highlighted in the main text, it is convenient to define the BCS ground-state in the pseudospin representation via the single-particle spin expectation values,
\begin{equation}
    \langle \hat{\sigma}^+_j \rangle_{\mathrm{gs}} = \frac{1}{2}\frac{\Delta_{\mathrm{gs}}}{\sqrt{\Delta_{\mathrm{gs}}^2 + \varepsilon_j^2}}, \quad \langle \hat{\sigma}^z_j \rangle_{\mathrm{gs}} = \frac{\varepsilon_j}{\sqrt{\Delta_{\mathrm{gs}}^2 + \varepsilon_j^2}} , \label{eqn:SM_BCS_gs_spin}
\end{equation}
where we have adopted the subscript $j$ to mirror the conventions of the cavity-QED case for simplicity. The crucial feature of this ground-state in terms of the dynamical phase diagram is that the sign of the initial inversion $\langle \hat{\sigma}^z_j \rangle_{\mathrm{gs}}$ is correlated with the sign of the single-particle energy splitting $\varepsilon_j$. It is this feature that we seek to mimic with our initial state, by correlating the sign of the azimuthal angle $\pm\Delta\phi_0/2$ of each spin ensemble with the sign of the energy splitting $\epsilon_0$. Based on our results and comparison with the literature, this type of correlation appears to be a necessary requirement for the observation of phase III. 

For completeness, and mindful of the rapidly advancing technical capabilities in state-of-the-art cavity-QED experiments, we also summarize in this SM the dynamical phase diagram which might be accessed with an initial state that more closely follows the BCS ground-state: $\vert \psi^{\prime}_0 \rangle = \vert \frac{\pi}{2} + \frac{\Delta\theta_0}{2},0\rangle_+ \otimes \vert \frac{\pi}{2} -\frac{\Delta\theta_0}{2},0\rangle_-$. Here, the relative opening angle $\Delta\theta_0$ between the ensembles is with respect to the elevation, which means that the signs of $\langle \hat{\sigma}^z_j \rangle$ will be correlated with the sign of the energy splitting $\epsilon_0$ in much closer correspondence to the BCS ground-state Eq.~(\ref{eqn:SM_BCS_gs_spin}), as can be seen from the initial expectation values $\vec{\sigma}_j(0) = (\mathrm{cos}(\Delta\theta_0/2),0,\pm\mathrm{sin}(\Delta\theta_0/2))$. In Fig.~\ref{fig:AnalyticRoots} we present the numerically evaluated dynamical phase diagram for this state and compare to the equivalent phase diagram for $\vert \psi_0 \rangle$. To be consistent with Fig.~2b of the main text we fix $\epsilon_0/(\chi N) = 0.1$ and probe the dependence on $W/(\chi N)$ and opening angle $\Delta\theta_0$. We limit our scan over the latter to $0 \leq \Delta\theta_0 \lessapprox \pi - \pi/8$ because our numerical algorithm performs poorly when the initial state becomes too closely aligned with the poles. We find that the dynamical phase diagram of the BCS-like state $\vert \psi^{\prime}_0 \rangle$ is very similar to the results in the main text, both qualitatively in terms of the presence of all three dynamical phases but also quantitatively in terms of the transition points as a function of $W/(\chi N)$ and $\Delta\theta_0$. 

We can make the last connection quantitatively rigorous by solving for the roots of the squared Lax vector in the same limiting cases ii) and iii) from our prior analysis. First, for $W\ll \epsilon_0, \chi N$ [case ii)] we again find two pairs of complex roots (excluding when $\Delta\theta_0 = \pm\pi$).
\begin{widetext}
\begin{equation}
    \begin{gathered}
    u_{1\pm} = -\frac{1}{4}\sqrt{4\epsilon_0^2 - 2\chi^2N^2 - 2\chi N\left[ \chi N \mathrm{cos}(\Delta\theta_0) - 4\epsilon_0\mathrm{sin}\left(\frac{\Delta\theta_0}{2}\right) \right]} \pm \frac{i\chi N}{2}\mathrm{cos}\left( \frac{\Delta\theta_0}{2} \right) , \\
    u_{2\pm} = \frac{1}{4}\sqrt{4\epsilon_0^2 - 2\chi^2N^2 - 2\chi N\left[ \chi N \mathrm{cos}(\Delta\theta_0) - 4\epsilon_0\mathrm{sin}\left(\frac{\Delta\theta_0}{2}\right) \right]} \pm \frac{i\chi N}{2}\mathrm{cos}\left( \frac{\Delta\theta_0}{2} \right) . \label{eqn:RootsSmallW}
    \end{gathered}
\end{equation}
\end{widetext}
This result indicates we should expect phase III for small enough inhomogeneity and regardless of the initial opening angle (excluding the trivial case when $\Delta\theta_0 = \pm\pi$). Secondly, for $\epsilon_0 \ll W/(\chi N)$ [case iii)] we have 
\begin{equation}
    u_{\pm} = \pm\frac{iW}{4} \mathrm{tan}\left[ \frac{W}{4\chi N}\mathrm{sec}\left(\frac{\Delta\theta_0}{2}\right)\right] ,
\end{equation}
for $W/(\chi N) < 2\pi \mathrm{cos}(\Delta\theta_0/2)$. This result demonstrates there exists a transition between phases II and I at $W/(\chi N) = 2\pi \mathrm{cos}(\Delta\theta_0/2)$ identically to our previous analysis of case iii) for the state $\vert \psi_0 \rangle$. 


\begin{figure}[!]
    \includegraphics[width=8cm]{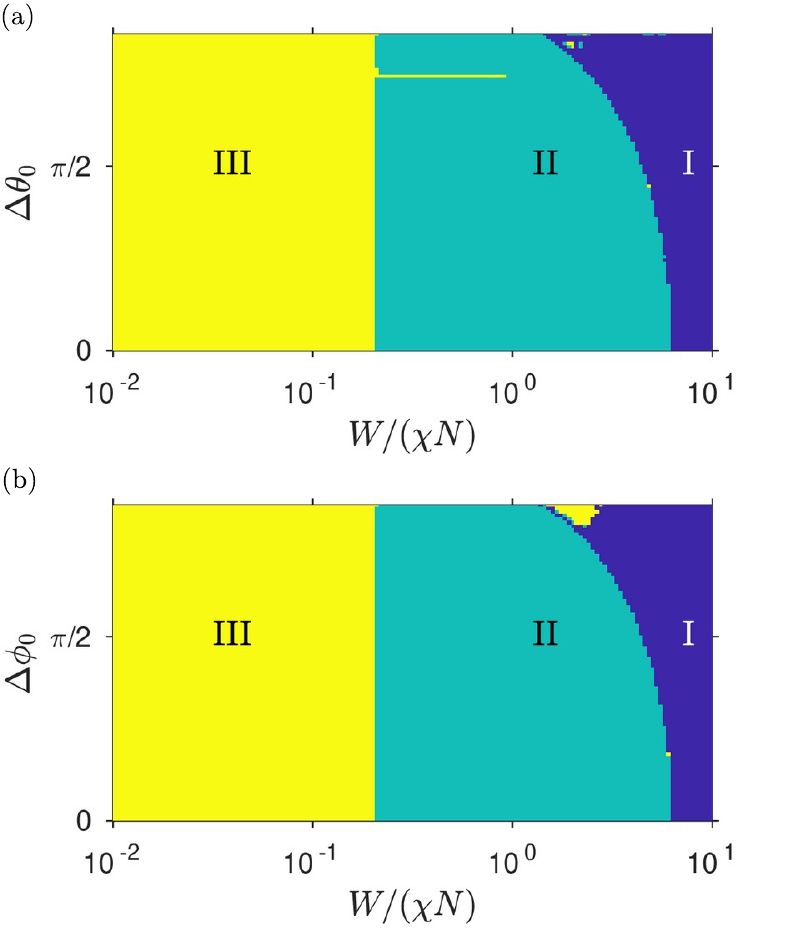}
    \caption{Dynamical phase diagram for: (a) $\vert \psi_0^{\prime} \rangle = \vert \pi/2+\Delta\theta_0/2,0\rangle_+ \otimes \vert \pi/2 -\Delta\theta_0/2,0\rangle_-$ and (b) $\vert \psi_0 \rangle = \vert \pi/2, \Delta\phi_0/2\rangle_+ \otimes \vert \pi/2, -\Delta\phi_0/2\rangle_-$. Parameters are identical to Fig.~2b of main text, particularly $\epsilon_0/(\chi N) = 0.1$. Some small structure, particularly the protrusion of phase III into phase II in panel (a), is likely a numerical artefact related to the breakdown of our numerical algorithm as it requires a minimum tolerance as an input to distinguish roots.}
    \label{fig:AnalyticRoots}
\end{figure}

\subsection{Analytic solution of phase III from Lax analysis}
The dynamics of the BCS pairing amplitude, $\vert \Delta(t) \vert$, in phase III can be solved exactly in certain cases. Specifically, by adopting an amplitude-phase parametrization of the pairing term, $\Delta(t) = \sqrt{R(t)}e^{i\varphi(t)}$, and substitution of this into the classical equations of motion \cite{Enolskii_2005,Levitov_2006} a description of the dynamics of the amplitude $R(t)$ can be reduced to the single differential equation
\begin{equation}
    \dot{R}^2 = 4(R_+ - R)(R-R_-)(R+\tilde{R}) . \label{eqn:Reqn}
\end{equation}
Here, $R_{\pm}$ and $\tilde{R}$ are obtained from the roots of the Lax vector \cite{Foster_2015,FosterPrivateComm}. Denoting the two pairs of complex roots as $\bar{u}_{1\pm} = \bar{u}_{1r} \pm i\bar{u}_{1i}$ and $\bar{u}_{2\pm} = \bar{u}_{2r} \pm i\bar{u}_{2i}$ we have $R_{\pm} = (\vert \bar{u}_{1i} \vert \pm \vert \bar{u}_{2i} \vert )^2$ and $\tilde{R} = (\bar{u}_{1r} - \bar{u}_{2r})^2$. Before proceeding further, we point out that some features of the pairing amplitude oscillations in phase III can already be predicted. Specifically, we have that the oscillation amplitude is $\mathcal{A} = \mathrm{max}(\vert\Delta(t)\vert) - \mathrm{min}(\vert \Delta(t) \vert) = \sqrt{R_+} - \sqrt{R_-}$. 

For certain cases, Eq.~(\ref{eqn:Reqn}) can be solved analytically. In particular, when $\Delta\phi_0 = \pi$ we have that $\vert \bar{u}_{1i} \vert = \vert \bar{u}_{2i}\vert$ [see Eq.~(\ref{eqn:RootsBB})] so that $R_- = 0$. Then, a solution to Eq.~(\ref{eqn:Reqn}) is
\begin{equation}
    \vert \Delta(t) \vert = \sqrt{R_+} \left\vert \mathrm{sn}\left( t\sqrt{\tilde{R}}, -\frac{R_+}{\tilde{R}} \right) \right\vert , \label{eqn:DeltaDynamics}
\end{equation}
where $\mathrm{sn}(u,m)$ is a Jacobi elliptic function. The nonlinear oscillations of $\vert \Delta(t) \vert$ can then be determined to have an amplitude $\mathcal{A} = \sqrt{R_+}$ and a period $T = 4K(-R_+/\tilde{R})/\sqrt{\tilde{R}}$ where $K(m)$ is the complete elliptic integral of the first kind. In the limit of $W \ll \epsilon_0 \ll \chi N$ we can simplify $R_+ = \tilde{R} \approx \epsilon_0\chi N/2$ from Eq.~(\ref{eqn:RootsBB}). This results in a pair of observations: i) the amplitude of oscillations vanishes as $\mathcal{A} = \sqrt{\epsilon_0\chi N}/\sqrt{2}$ for small splitting, and similarly ii) the frequency $\omega_{\mathrm{osc}}$ of oscillations is approximately $\omega_{\mathrm{osc}} \sim 1/T \propto \sqrt{\epsilon_0 \chi N}$. This latter result is clearly illustrated in Fig.~3d of the main text. 

\begin{figure}[!]
    \includegraphics[width=8cm]{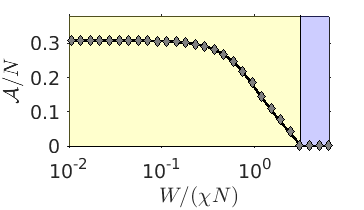}
    \caption{Amplitude $\mathcal{A} \equiv \mathrm{max}(\vert \Delta(t) \vert) - \mathrm{min}(\vert \Delta(t) \vert)$ of oscillations in $\vert \Delta(t) \vert$ obtained analytically from Eqs.~(\ref{eqn:DeltaDynamics}) and (\ref{eqn:RootsBB}) (solid black line) and numerical integration of mean-field equations of motion following from the Hamiltonian (\ref{eqn:BCSham}) for $N=10^3$ (grey markers). Calculations are for an initial state with $\Delta\phi = \pi$, $\epsilon_0/(\chi N) = 0.1$ and varying $W/(\chi N)$. Phases I and III, equivalent to $\mathcal{A} = 0$ or $\mathcal{A} \neq 0$ in this case, are indicated by the blue and yellow backgrounds, respectively.}
    \label{fig:Amplitude_CompareMethods}
\end{figure}

As further verification of both our Lax analysis and the predictions of Eq.~(\ref{eqn:DeltaDynamics}) we compare the amplitude $\mathcal{A}$ for dynamics of the initial state with $\Delta\phi_0 = \pi$ obtained from: i) the analytic solution of the Lax roots Eq.~(\ref{eqn:RootsBB}), and ii) full numeric integration of the mean-field equations of motion following from the Hamiltonian (\ref{eqn:BCSham}). Results are plotted in Fig.~\ref{fig:Amplitude_CompareMethods}. We observe excellent quantitative agreement, not only in terms of the boundary between phases I and III but also the predicted $\mathcal{A}$ as a function of $W/(\chi N)$. Small differences are entirely attributable to finite size effects (e.g., analytic results assume $N\to\infty$ and thus a \emph{continuous} distribution of $\varepsilon_j$). 

\subsection{Phases IIIa and IIIb}
We also report in the main text an identification of sub-phases IIIa and IIIb in the limit $W \ll \chi N, \epsilon_0$. Our classification of these sub-phases is instead related to the characterization of the real and imaginary parts of the two pairs of complex roots present for phase III. 

To be concrete, we define sub-phases IIIa and IIIb by the effective order parameter $R_-$, which was previously introduced in the differential equation (\ref{eqn:Reqn}). To recap, $R_- = (\vert \bar{u}_{1i} \vert - \vert \bar{u}_{2i} \vert )^2$ where we have written the roots of the squared Lax vector in the form $\bar{u}_{1\pm} = \bar{u}_{1r} \pm i\bar{u}_{1i}$ and $\bar{u}_{2\pm} = \bar{u}_{2r} \pm i\bar{u}_{2i}$. Phase IIIa exists for $R_- = 0$ and phase IIIb for $R_- \neq 0$. Inspecting Eq.~(\ref{eqn:Reqn}) we observe that $R_{\pm}$ define the maximum and minimum values of the oscillations in $\vert \Delta(t) \vert^2$, so clearly we can also interpret phase IIIa as oscillations where the pairing amplitude periodically vanishes $\vert \Delta \vert = 0$, whereas in phase IIIb the pairing amplitude is strictly greater than zero always, $\vert \Delta \vert > 0$. 

Mathematically, the condition $R_- = 0$ corresponds to the case where the magnitude of the imaginary part of the Lax roots is identical for both pairs, i.e., $\vert \bar{u}_{1i} \vert = \vert \bar{u}_{2i} \vert$. In the limit of $W \ll \chi N, \epsilon_0$ we can use the results of Eq.~(\ref{eqn:Roots_SmallW}) to identify that this condition occurs for a critical splitting
\begin{equation}
    \epsilon_0^{\mathrm{c}} = \frac{\chi N}{2}[1 + \mathrm{cos}(\Delta\phi_0)] .
\end{equation}

\section{Experimentally realistic model \label{sec:ExperimentalModel}}
In Fig.~4 of the main text we present quantitative predictions illustrative of the dynamical phases based upon state-of-the-art cavity-QED experiments. Here, we summarize the model these calculations are based upon. For the interested reader, further detail regarding the derivation of our model can be found in Ref.~\cite{Muniz2020} and the associated Supplementary Information. 

\subsection{Emulated BCS-like Hamiltonian}
We consider an ensemble of $2N$ atoms confined in a standing wave optical lattice supported by a cavity. Each atom encodes a spin-$1/2$ in a pair of long-lived electronic states, $\vert \uparrow \rangle$ and $\vert \downarrow \rangle$, which are separated by a narrow linewidth optical transition of frequency $\omega_a$. A single common cavity mode couples the electronic states with single-photon Rabi frequency $2g$. In the limit that the cavity mode at frequency $\omega_c$ is far detuned from the atomic transition $\vert \delta_c \vert = \vert\omega_c - \omega_a\vert \gg g\sqrt{N}$, then the cavity field can be adiabatically eliminated and the intracavity photons serve only to mediate effective exchange interactions between the spins. Single-particle energy shifts can be generated by applying external fields to generate Stark or Zeeman shifts of the electronic states. Combining these effects with relevant sources of decoherence leads to an effective description for the atomic system in terms of a Lindblad master equation for the density matrix $\hat{\rho}_a$,
\begin{equation}
    \frac{d\hat{\rho}_a}{dt} = -\frac{i}{\hbar}\left[ \hat{H}, \hat{\rho}_a \right] + \mathcal{L}_s[\hat{\rho}_a] , \label{eqn:MasterEqn}
\end{equation}
with Hamiltonian
\begin{equation}
    \hat{H} = \hbar\sum_{i,j} \chi_{ij} \hat{\sigma}^+_i \hat{\sigma}^-_j + \sum_i \varepsilon_j\hat{\sigma}^z_i , \label{eqn:InhomHamiltonian}
\end{equation}
and decoherence due to spontaneous emission at rate $\gamma_s$ described by the Lindblad jump operator,
\begin{equation}
    \mathcal{L}_s[\hat{\rho}] = \frac{\gamma}{2}\sum_i 2\hat{\sigma}^-_i\hat{\rho}\hat{\sigma}^+_i - \hat{\sigma}^+_i\hat{\sigma}^-_i\hat{\rho} - \hat{\rho}\hat{\sigma}^+_i\hat{\sigma}^-_i  .
\end{equation}
In the Hamiltonian (\ref{eqn:InhomHamiltonian}) the all-to-all interactions are characterized by $\chi_{ij} = -g_ig_j/\delta_c$. The inhomogeneity is inherited from the spatial variation of the atom-light coupling, $g_j \propto g\mathrm{cos}(k_d j)$ for $k_d = \pi \lambda_L/\lambda_c$, due to the incommensurate wavelengths of the confining lattice, $\lambda_L = 813$~nm, and cavity mode, $\lambda_c = 689$~nm. In numerical simulations (see later discussion) we typically observe that the main consequence of inhomogeneous interactions is to effectively re-scale the interactions to their root-mean-square value, $\chi \to \chi/2$. This is also consistent with related work published in Refs.~\cite{Norcia_2018,Muniz2020}. 

\subsection{State preparation}
State preparation is realized by a combination of coherent single-particle rotations and shifts of the internal atomic levels. Specifically, we assume that all spins are initially prepared in the single-particle state $\vert \downarrow \rangle_j$. The cavity is suddenly injected with coherent light which is tuned to be resonant with the atomic transition \cite{Muniz2020}. This process can be modelled as a single-particle term $H_{\mathrm{rot}} = \sum_j\frac{\Omega_j}{2}\hat{\sigma}^y_j$. The inhomogeneity of the driving term $\Omega_j = \Omega_0\mathrm{cos}(k_d j) \propto g_j$ is again inherited from the inhomogeneous atom-light coupling $g_j$. In our simulations, we apply $H_{\mathrm{rot}}$ for a time $\tau_{\mathrm{rot}}$ so that $\Omega_0\tau_{\mathrm{rot}} = \pi/2$, i.e., a $\pi/2$-pulse is engineered with respect to the strongest coupled atoms, i.e., $\Omega_j = \Omega_0$. For simplicity, we assume $\Omega_0 \gg \chi N, \epsilon_0, W$ such that interactions and energy shifts can be ignored during state preparation. 

After the single-particle rotation sequence, the variable opening angle $\Delta\phi_0$ of the initial state can be generated by suddenly turning on a (selective) large uniform energy splitting between the atoms of each ensemble, e.g., $\varepsilon_j = \pm \epsilon_0$, to generate precession of each ensemble about the $z$-axis. This could be achieved, for example, by a spatially selective optical Stark shift of the internal levels $\vert \uparrow \rangle$ and $\vert \downarrow \rangle$. 

The combination of these two coherent operations leads to initial states of the form $\vert \psi_0 \rangle = \vert \psi_0\rangle_+ \otimes \vert \psi_0 \rangle_-$ where $\vert \psi_0 \rangle_{\pm} = \bigotimes_{j} \left[ \mathrm{cos}(\theta_j/2)\vert \downarrow \rangle_j + e^{ \pm i\Delta\phi/2}\mathrm{sin}(\theta_j/2) \vert \uparrow \rangle_j \right]$ with $\theta_j = (\pi/2)\mathrm{cos}(k_d j)$ and $j$ runs over $1,...,N$ or $N+1,...,2N$ for the respective ensembles.

\subsection{Measurement of pairing amplitude via leaked light}
The BCS pairing amplitude, proportional to the transverse spin coherence, can be monitored in experiment by detecting the light which leaks out through the cavity mirrors. Specifically, when the cavity field is eliminated to yield the effective spin model Eq.~(\ref{eqn:MasterEqn}), at the mean-field level the intra-cavity field is related to the spins via
\begin{equation}
    \langle \hat{a} \rangle = \frac{-2}{2\delta_c - i\kappa}\sum_j g_j \langle\hat{\sigma}^-_j\rangle .
\end{equation}
We use this relation to plot the dynamics of the intracavity field in Fig.~4 of the main text.

\subsection{Numerical simulation and parameters}
To simulate the experimental system we solve the mean-field equations of motion associated with the master equation (\ref{eqn:MasterEqn}). Specifically, we use the software package XMDS2 \cite{XMDS2} to numerically integrate the system of equations generated by $\dot{\mathcal{O}} = \mathrm{Tr}[\hat{\mathcal{O}} \frac{d}{dt}\hat{\rho}_a]$ with the approximation $\langle\hat{\mathcal{O}}_1(t)\hat{\mathcal{{O}}}_2(t)\rangle=\langle\hat{\mathcal{O}}_1(t)\rangle \langle\hat{\mathcal{{O}}}_2(t)\rangle$. 

To account for inhomogeneities in the effective spin-spin interactions and state preparation we model a system composed of two distinct ensembles ($\pm$) of atoms spatially distributed at lattice sites $j = 1,2,...,2N_{\mathrm{sim}}$. Typically, we take $N_{\mathrm{sim}} = 10^2-10^3$ and re-scale all parameters and results to match a true atom number of $N=10^6$. We adopt relevant parameters from Ref.~\cite{Muniz2020}: $g/(2\pi) = 10.9$~kHz, $\gamma_s/(2\pi) = 7.5$~kHz, $\delta_c/(2\pi) = -50$~MHz and $\kappa/(2\pi) = 153$~kHz.


\end{document}